\documentclass[preprint, 12pt]{revtex4-1}

\usepackage{amsmath,graphicx,bbm,mathrsfs,amssymb,pst-all,bm,color}
\usepackage{subfigure}

\begin{document}

\title{Saving resources through repeat-until-success positive-operator-valued-measure measurements in quantum computation}
\author{Hefeng Wang$^1$}
\email{wanghf@mail.xjtu.edu.cn}
\author{Sixia Yu$^2$}
\email{yusixia@ustc.edu.cn}
\author{Hua Xiang$^3$}
\email{hxiang@whu.edu.cn}
\affiliation{$^{1}$School of Physics, Xi'an Jiaotong University, Xi'an,
710049, China}
\affiliation{$^{2}$Hefei National Laboratory for Physical Sciences at
Microscale and Department of Modern Physics, University of Science and
Technology of China, Hefei, Anhui 230026, China}
\affiliation{$^{3}$School
of Mathematics and Statistics, Wuhan University, Wuhan, 430072, China}

\begin{abstract}
We present a quantum computation approach in which computation is guided by
positive-operator-valued-measure~(POVM) measurements following a given
computation path in multisteps. In this approach, one ancillary qubit is
coupled to a register of working qubits, and a POVM measurement is
implemented effectively on the working qubits by applying a unitary
operation on the whole system followed by a projective measurement performed
on the ancillary qubit. Each step of the computation is a
repeat-until-success procedure such that the desired state of the step is
obtained deterministically on the working qubits via POVM measurements. The
principle of deferred measurement states that measurements can always be
moved from an intermediate stage of a quantum circuit to the end of the
circuit without affecting the efficiency of the computation. We demonstrate
that in our approach, by introducing intermediate measurements on the
ancillary qubit in the computation process, both the number of qubits and
unitary operations can be reduced polynomially, compared to the case where
the intermediate measurements are deferred to the end of the computation. We
also provide a method for implementation of the approach.
\end{abstract}

\maketitle

\section{Introduction}

Saving resources is important in quantum computation, given the difficulty
of integrating a large number of qubits and their limited coherence time.
Designing a quantum algorithm that uses fewer qubits and quantum gates to
solve a problem is therefore an important subject in the field of quantum
computation. Various methods have been explored to reduce the resources in
quantum computation. For instance, the quantum signal processing method
achieves the optimal scaling in gate count for Hamiltonian simulation~\cite%
{QSP}. Quantum measurement has been used to enhance the computational power
of unitary circuits at a lower cost in qubit and circuit resources~\cite%
{mea, bro, sha, poulin}. In Ref.~\cite{parker}, it was shown that only one
ancillary qubit is needed to implement Shor's algorithm by introducing
intermediate measurements in the computation, and this method was
implemented on an ion-trap system~\cite{iontrap}. Furthermore, by adopting
probabilistic quantum circuits known as repeat-until-success~(RUS) circuits~%
\cite{wiebe,pae,boc,dong}, in which quantum measurement is used to herald
the successful application of a desired unitary operation, the resources
required to implement unitary operations on a quantum computer can be
reduced dramatically.

In quantum computation, instead of evolving from an initial state of the
system to the final state directly, it can be more efficient to follow a
computation path by evolving from the initial state to the final state
through a sequence of intermediate states step by step, e.g., the jagged
adiabatic path~\cite{aharanov}, and the multistep nonlinear adiabatic
evolution path~\cite{sch} in which the first-order quantum phase transition
can be avoided in state preparation. In this work, we introduce a quantum
computation approach that follows a given computation path step by step,
with each step being implemented via a new RUS procedure. This approach
significantly reduces both the number of qubits and unitary operations,
while simplifying complexity of the quantum circuit.

We consider a quantum computation process following a state evolution path
\begin{equation}
|\varphi ^{(0)}\rangle \rightarrow |\varphi ^{(1)}\rangle \rightarrow \cdots
\rightarrow |\varphi ^{(m)}\rangle ,
\end{equation}%
in which a step $|\varphi ^{(k-1)}\rangle \rightarrow |\varphi ^{(k)}\rangle
$ of the computation is realized through a new RUS procedure that
effectively performs positive-operator-valued-measure~(POVM)
measurements~resulting in state $|\varphi ^{(k)}\rangle $~($k=1,\ldots ,m$).
By coupling one ancillary qubit to a register of working qubits, a POVM
measurement on the working qubits is realized effectively by applying a
unitary operation on the whole system and followed by a projective
measurement performed on the ancillary qubit. A RUS procedure is used
to obtain the desired state of the step deterministically on the working
qubits via a number of POVM measurements.

In the general RUS protocol, a unitary operator acts on ($n+l$) qubits, with
$n$ working qubits and $l$ ancillary qubits, followed by a measurement
performed on the ancillary qubits, one measurement outcome is labeled
\textquotedblleft success\textquotedblright\ and other measurement outcomes
are labeled \textquotedblleft failure\textquotedblright. When a failure
measurement is observed, a unitary operation is performed to transform the
working qubits back to their original input state. This procedure is
repeated until a success measurement is observed. The RUS procedure can be
equated to a coherent gate as a success measurement is certified. Here, we
propose a new RUS protocol that consists of ($n+1$) qubits with one
ancillary qubit and $n$ working qubits, which is characterized by three
features: ($i$) Quantum entanglement between the ancillary qubit and the
working qubits is introduced explicitly in the protocol; ($ii$) A POVM
measurement is implemented on the working qubits via a unitary operation on
the whole system followed by a projective measurement on the ancillary
qubit, yielding either the initial state or the desired state of the RUS
procedure; ($iii$) A series of RUS procedures can be concatenated
sequentially to realize a RUS-POVM computation process. In this RUS
protocol, a unitary operation is applied on the qubits to generate an
entangled state of the ancillary qubit and the working qubits. Then a
projective measurement is performed on the ancillary qubit, and the
measurement outcome tells whether the desired state of the RUS procedure is
obtained or not on the working qubits. In the case where a failure
measurement occurs, the working qubits remain in their initial state, and
the RUS procedure is repeated until a success measurement is observed, and
the working qubits are in the desired state. Each iteration of applying a
unitary operation followed by a measurement on the ancillary qubit
effectively implements a POVM measurement on the working qubits, either
producing the desired state of the step or staying in the initial state.

The principle of deferred measurement states that measurements can always be
moved from an intermediate stage of a quantum circuit to the end of the
circuit; if the measurement results are used at any stage of the circuit,
then the classically controlled operations can be replaced by conditional
quantum operations, without affecting efficiency of the computation~\cite{nc}%
. We show that by performing measurements in the intermediate stages of the
computation, the RUS-POVM approach achieves polynomial reduction in quantum
resources, including both the number of qubits and unitary operations, while
dramatically simplifying circuit complexity, as compared to the case where
intermediate measurements are deferred to the end of the circuit.

The structure of this work is as follows: In Sec.~II, we present the
RUS-POVM approach for quantum computation; in Sec.~III, we provide a method for
implementing this approach through quantum resonant transitions~(QRTs);
in Sec.~IV, we perform error analysis of implementing the RUS-POVM approach; in
Sec.~V, we compare the RUS-POVM approach with other RUS protocols and
measurement-assisted methods; and we close with a conclusion section.

\section{RUS-POVM approach}

In the RUS-POVM approach for quantum computation, the system contains one
ancillary qubit and an $n$-qubit register $R$ representing the working
qubits. For a computation where the register $R$ evolves following the state
evolution path in Eq.~($1$), the RUS-POVM approach can be described as
\begin{equation}
R_{m}\cdots R_{2}R_{1}|\varphi ^{(0)}\rangle ,
\end{equation}%
where $R_{k}=A_{1\mid k}A_{0\mid k}^{n_{k}-1}$ represents the $k$th step of
the computation in which the RUS procedure succeeds in $n_{k}$~(which is not
certain) iterations, and $A_{0\mid k}$ and $A_{1\mid k}$ are effective
measurements associated with POVM elements $\left\{ A_{0\mid k}^{\dagger
}A_{0\mid k},A_{1\mid k}^{\dagger }A_{1\mid k}\right\} $. In the $k$th step,
a unitary operation $U_{k}$ is applied on the initial state $|0\rangle
|\varphi ^{(k-1)}\rangle $ of the step, followed by a projective measurement
on the ancillary qubit. This process can be described by operators $A_{0\mid
k}=\langle 0|U_{k}|0\rangle _{a}$ and $A_{1\mid k}=\langle 1|U_{k}|0\rangle
_{a}$, respectively. The unitary operation $U_{k}$ acts on the initial state
$|0\rangle |\varphi ^{(k-1)}\rangle $ to generate an entangled state%
\begin{equation}
U_{k}|0\rangle |\varphi ^{(k-1)}\rangle =a_{k}|0\rangle |\varphi
^{(k-1)}\rangle +b_{k}|1\rangle |\varphi ^{(k)}\rangle ,
\end{equation}%
where $\left\vert a_{k}\right\vert ^{2}+\left\vert b_{k}\right\vert ^{2}=1$,
and $\left\vert b_{k}\right\vert $ is polynomial large. The operator $%
A_{0\mid k}$ represents that measurement outcome on the ancilla is in state $%
|0\rangle $, while $A_{1\mid k}$ means measurement outcome on the ancilla is
in state $|1\rangle $. The POVM measurement results in either state $%
|\varphi ^{(k-1)}\rangle $ or state $|\varphi ^{(k)}\rangle $ on the
register $R$. Fig.~$1$ shows a RUS-POVM computation process that contains
$m$ RUS procedures.

In the $k$th step of the RUS-POVM approach, by protecting state $|\varphi
^{\left( k-1\right) }\rangle $ in the entangled state of Eq.~($3$), this
state can be used repeatedly for obtaining the desired state $|\varphi
^{\left( k\right) }\rangle $ of the step deterministically. Here
\textquotedblleft deterministically\textquotedblright\ means that by running
procedures of a step repeatedly, we know definitely when the desired state
of the step is obtained. If the measurement outcome is in state $|1\rangle $%
, it indicates a success measurement and the desired state $|\varphi
^{\left( k\right)}\rangle $ is obtained on $R$, then we run the ($k+1$)th
step; otherwise if the ancilla is in state $|0\rangle $, it means that the
register $R$ remains in state $|\varphi ^{\left(k-1\right) }\rangle $, then
we repeat procedures of the step until a success measurement is observed.
The $k$th step of the computation is a RUS procedure that effectively
performs POVM measurements yielding state $|\varphi ^{(k-1)}\rangle $ or $%
|\varphi ^{(k)}\rangle $ on the register $R$. The number of times $n_{k}$ of
procedures of the $k$th step has to be repeated is proportional to $1/p_{k}$%
, where $p_{k}=\left\vert b_{k}\right\vert ^{2}$ is the success probability
of the measurement on the ancilla in the $k$th step. Therefore, the
runtime of the computation is proportional to $\sum_{k=1}^{m}1/p_{k}$, which
scales linearly with the number of steps in the computation as long as $%
p_{k} $ is polynomial large.
\begin{figure}[tbp]
\centering
\includegraphics[width=0.98\columnwidth, clip]{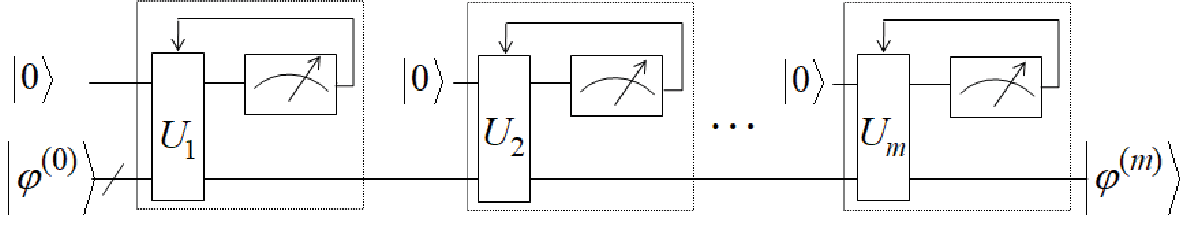}
\caption{Quantum circuit for the RUS-POVM computation process. The circuit in
the dashed square can be run repeatedly.}
\end{figure}
\begin{figure}[bp]
\centering
\includegraphics[width=0.98\columnwidth, clip]{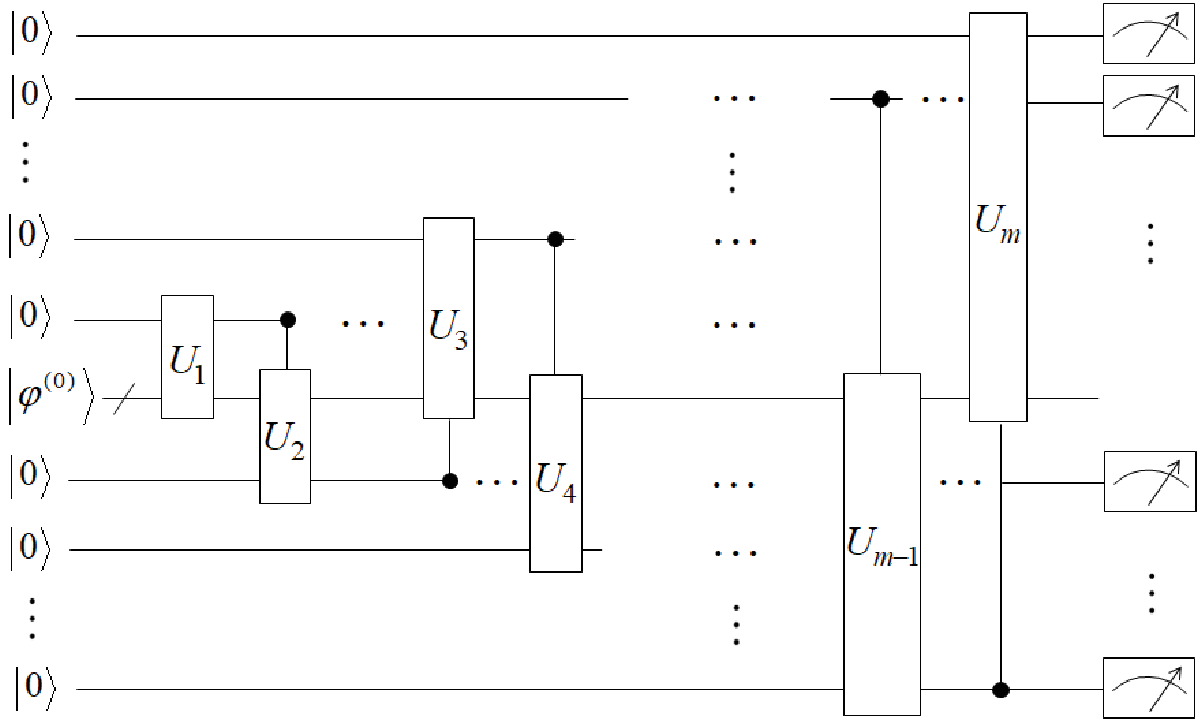}
\caption{Quantum circuit by deferring the intermediate measurements in the
RUS-POVM approach of Fig.~$1$ to the end of the circuit, and replacing the
intermediate measurements by controlled unitary operations.}
\end{figure}

If the intermediate measurements on the ancillary qubit in the RUS-POVM
approach are simply moved to the end of the circuit and replaced by
controlled unitary operations, following the principle of deferred
measurement, it becomes a quantum computation with $m$ controlled qubits as
shown in Fig.~$2$. This computation succeeds only when the measurement
outcomes on all the control qubits are in state $|1\rangle $ simultaneously,
the probability of which is $\prod\limits_{k=1}^{m}p_{k}$. Therefore the
runtime of a successful computation in Fig.~$2$ is proportional to $%
\prod\limits_{k=1}^{m}1/p_{k}$, which is not efficient.

Usually measurements in the intermediate stages of a computation are
performed only once, there is no success or failure measurement. For
example, it was shown in Refs.~\cite{parker,iontrap} that by introducing
intermediate measurements in the quantum circuit, the phase estimation
algorithm can be implemented by using only one ancillary qubit, where an
intermediate measurement is performed to reveal the state of the ancilla,
which is either $|0\rangle $ or $|1\rangle $ with certainty. The measurement
outcome is used to decide whether or not to perform a rotation in the
quantum Fourier transform. In the RUS-POVM approach, however, the
measurement outcome on the ancilla is uncertain, only one measurement
outcome is defined as success measurement and the other outcome is failure.
The procedures of the step are performed repeatedly until a success
measurement is observed. If the intermediate measurements are deferred to
the end of the circuit, a success computation means that all the measurement
outcomes on the control qubits must be success measurements, the probability
of which decreases exponentially with respect to the number of steps in the
computation.
\begin{figure}[tbp]
\centering
\includegraphics[width=0.98\columnwidth, clip]{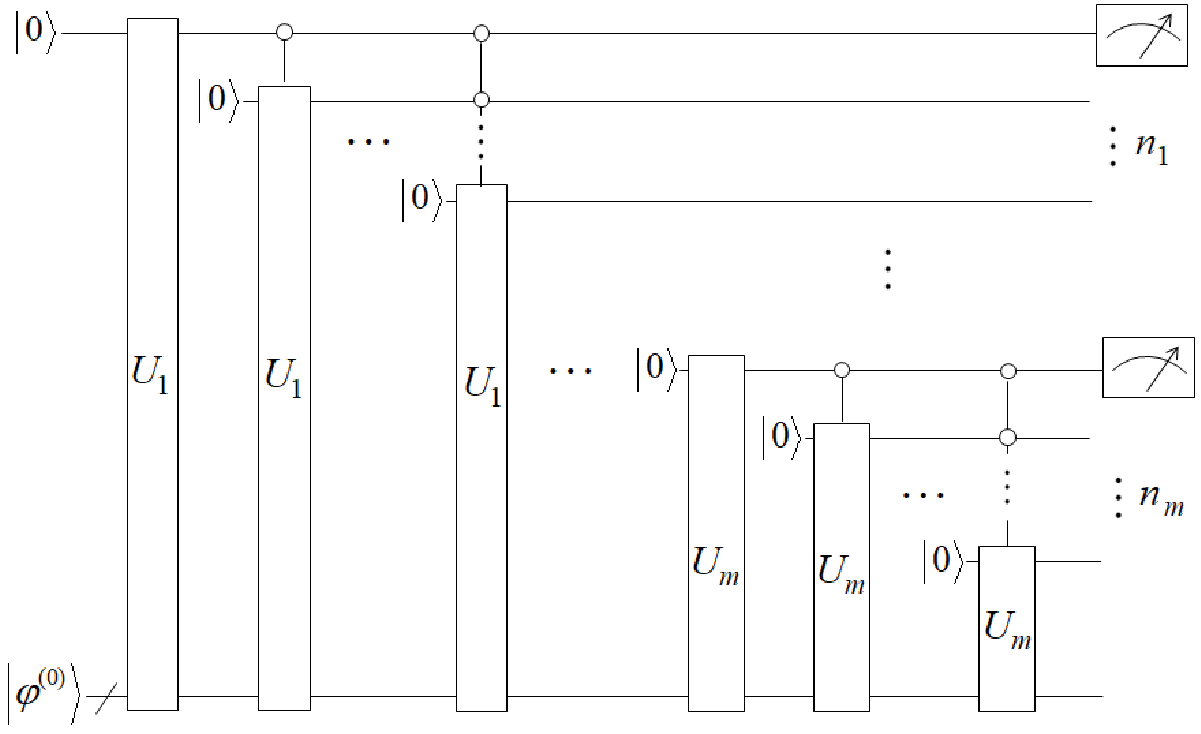}
\caption{Quantum circuit for simulating the RUS-POVM computation process of
Fig.~$1$ by performing measurements at the end of the circuit.}
\end{figure}

Each step of the RUS-POVM approach is a probabilistic procedure, the number
of times of applying a unitary operation followed by a projective
measurement on the ancilla is not certain. Whether or not to proceed to the $%
(k+1)$th step depends on the result of the $k$th step heralded by
measurement outcome on the ancilla, and the desired state of the $k$th step
is used as the initial state of the $(k+1)$th step. To simulate a step of
the RUS-POVM approach with coherent evolution, we have to consider all
possible measurement outcomes, i.e., by applying the unitary operation $%
U_{k} $ followed by a projective measurement on the ancilla for $n_{k}$
iterations, a success measurement may occur at any iteration, and a coherent
implementation of the step has to traverse all possible cases. Such a
circuit for simulating the RUS-POVM approach is shown in Fig.~$3$, it
simulates the RUS-POVM approach efficiently with all the intermediate
measurements deferred to the end of the circuit. This leads to more overheads
than that of the RUS-POVM approach in Fig.~$1$. A RUS procedure in the RUS-POVM approach
is simulated by applying a number of multi-qubit $|0\rangle $-controlled
unitary operations and performing measurements at the end of the circuit.
The number of ancillary qubits for simulating the $k$th step is $n_{k}$, and
the probability of observing at least one success measurement is $%
1-(1-p_{k})^{n_{k}}$. Therefore the success probability of the computation
in Fig.~$3$ is $\prod\limits_{k=1}^{m}\left[ 1-(1-p_{k})^{n_{k}}\right] $,
which is polynomial large for finite $p_{k}$. The circuit in Fig.~$3$ has
the same efficiency as that of the RUS-POVM approach in Fig.~$1$, but
requires more quantum resources. In the RUS-POVM approach, a
unitary operation $U_{k}$ is quried for $n_{k}$ times until a successful
measurement is obtained on the ancillary qubit. Here $n_{k}$ is not certain,
this is an advantage of the approach for simplifying the circuit. In the
circuit of Fig.~$3$ for simulating the RUS-POVM approach, $n_{k}$ has to be
fixed in order to construct the circuit, and this increases the circuit
complexity.

There is only one ancillary qubit and $\Sigma _{k=1}^{m}n_{k}$ unitary
operations in the RUS-POVM approach. The quantum circuit in Fig.~$3$
simulates the RUS-POVM approach efficiently with more resources. In the
circuit of Fig.~$3$, we need to implement a series of multi-qubit controlled
unitary operations to simulate a step of the RUS-POVM approach. For an $n$%
-qubit controlled unitary operation $C^{n}$-$U$, by introducing ($n-1$)
ancillary qubits, it can be implemented by using one single-qubit controlled
unitary operation $C$-$U$, and $2(n-1)$ Toffoli gates~\cite{barenco}. In
simulation of the $k$-th step of the RUS-POVM approach, we need $(n_{k}-1)$
more ancillary qubits, and these ancillas can be used in the following steps
of the approach. We also need one unitary operation $U_{k}$ and $n_{k}-1$
controlled unitary operations $C$-$U_{k}$ and $(n_{k}^{2}-n_{k})$ Toffoli
gates. Therefore we need a total number of $m $ unitary operations $U_{k}$ ($%
k=1,\ldots ,m$) and $\sum_{k=1}^{m}n_{k}-m$ single-qubit controlled unitary
operations, and $\sum_{k=1}^{m}(n_{k}^{2}-n_{k})$ Toffoli gates for
simulating the RUS-POVM approach by deferring all the intermediate
measurements to the end of the circuit. The total number of ancillary qubits
is $\sum_{k=1}^{m}n_{k}+n_{k}^{\max }-1$, where $n_{k}^{\max }$ is the
maximum of $n_{k}$.

\section{Implementation of the RUS-POVM approach}

The unitary operation generating the entangled state in Eq.~($3$) is the key
to the RUS-POVM approach, such that the intermediate calculation results of
each step can be protected and used repeatedly. If we can find a method to
realize such a unitary operation, then we can construct a RUS-POVM
computation. Here we introduce a method for constructing this unitary
operation based on a physical process of QRT~\cite%
{whf1,whf2,wy}.

If we can construct a Hamiltonian $H_{k}$ such that the state $|\varphi^{(k)}\rangle $ in the state evolution path of Eq.~($1$) is an eigenstate of $H_{k}$ with corresponding eigenvalue $E^{(k)}$, then the QRT method can be used for implementing the RUS-POVM approach. We first construct a Hamiltonian evolution path from an initial Hamiltonian $H_{0}$ to the problem Hamiltonian $H_{m}$ as: $H_{0}\rightarrow H_{1}\rightarrow \cdots
\rightarrow H_{m}$, then starting from the eigenstate $|\varphi^{(0)}\rangle $ of $H_{0}$, the system can be evolved through eigenstates of the intermediate Hamiltonians sequentially to reach the eigenstate $|\varphi^{(m)}\rangle $ of $H_{m}$ in $m$ steps. The unitary operation $U_{k}$ is constructed as time evolution of a Hamiltonian that bridges the Hamiltonians $H_{k-1}$ and $H_{k}$. It generates an entangled state as shown in Eq.~($3$) as the resonance condition is satisfied. There are different ways to
construct such a Hamiltonian evolution path, for example, we can select a number of discrete points on the adiabatic evolution path from $H_{0}$ to $H_{m}$. Moreover, for some problems, we can construct a sequence of Hamiltonians with increasing dimension to reach the final Hamiltonian as shown in Ref.~\cite{wy}.

In the following, we describe details of implementing the RUS-POVM approach through the QRT method. In the $k$th step of the approach, given the Hamiltonians $H_{k}$, $H_{k-1}$ and its eigenstate $|\varphi ^{\left(k-1\right) }\rangle $ and the corresponding eigenvalue $E^{\left( k-1\right)}$ obtained from the previous step, we are to evolve the system to the eigenstate $|\varphi ^{\left( k\right) }\rangle $ of $H_{k}$. For
simplicity, we assume that the corresponding eigenvalue $E^{\left( k\right)}$ of $H_{k}$ is known in advance. There is a method for obtaining $E^{\left(k\right) }$ by using the QRT method~\cite{wy}. The algorithm Hamiltonian of the $k$th step is constructed as
\begin{equation}
H^{\left( k\right) }=-\frac{1}{2}\omega \sigma _{z}\otimes
I_{N}+H_{R}^{\left( k\right) }+c\sigma _{x}\otimes B_{k},
\end{equation}%
where
\begin{equation}
H_{R}^{\left( k\right) }=\alpha _{k}|0\rangle \langle 0|\otimes
H_{k-1}+|1\rangle \langle 1|\otimes H_{k}\mathbf{,}\text{ \ \ \ }k=1,\cdots
,m,
\end{equation}%
and $B_{k}$ is an operator acting on the register $R$, here we set $B_{k}$
as $N$-dimensional identity operator $I_{N}$ for simplicity, and $\sigma
_{x} $, $\sigma _{z}$ are the Pauli matrices. The first term in Eq.~($4$) is
the Hamiltonian of the ancillary qubit, the second term contains Hamiltonian
of the register $R$ and describes the interaction between the ancillary
qubit and $R$, the third term is a perturbation. The parameter $\alpha _{k}$
is used to rescale the energy levels of $H_{k-1}$, and $c\ll 1$.

We set the parameter $\alpha _{k}$ such that $\alpha _{k}E^{\left(
k-1\right) }-E^{\left( k\right) }=\omega $, which is the condition of
quantum resonance between the ancillary qubit and the transition between
states $|\varphi ^{\left( k-1\right) }\rangle $ and $|\varphi ^{\left(
k\right) }\rangle $, and set the evolution time as $t_{k}$. Procedures of
the $k$th step of the computation are as follows:

$i$) Set the system in state $|0\rangle|\varphi^{\left( k-1\right)}\rangle $.

$ii$) Implement time evolution operator $U_{k}^{\prime }=\exp \left[
-iH^{\left( k\right) }t_{k}\right] $.

$iii$) Measure the ancillary qubit in its computational basis.

If the measurement outcome is in state $|1\rangle $, it means the state $%
|\varphi ^{\left( k\right) }\rangle $ is obtained, and we proceed to the ($%
k+1$)th step; otherwise if the measurement outcome stays in state $|0\rangle
$, it means that the register $R$ remains in state $|\varphi ^{\left(
k-1\right) }\rangle $, then we repeat procedures $ii$)-$iii$) until the
measurement outcome is $|1\rangle $. The above procedures can approximate
the RUS-POVM process in Eq.~($2$) within high accuracy as shown in the next
section. After applying the operator $U_{k}^{\prime }$, the system is
approximately in an entangled state $\sqrt{1-p_{k}}|0\rangle |\varphi
^{\left( k-1\right)}\rangle +\sqrt{p_{k}}|1\rangle |\varphi ^{\left(
k\right) }\rangle $, where $p_{k}\approx \sin ^{2}\left( ct_{k}d_{k}\right) $
is excitation probability of the ancilla that describes Rabi-oscillation
dynamics, in which the ancilla and the register $R$ exchange an excitation,
and $d_{k}=\left\vert \langle \varphi ^{(k-1)}|\varphi^{(k)}\rangle
\right\vert $. Therefore a RUS procedure is constructed via the QRT method.
The RUS-POVM computation can be run efficiently provided that $i$) the
overlap between two adjacent states of the state evolution path and, $ii$)
the energy gap between the eigenstate $|\varphi^{(k)}\rangle $ and its
nearest neighboring eigenstate of each Hamiltonian, are polynomial large.

We need to know the eigenvalue of an eigenstate of the intermediate
Hamiltonian to set the parameters to satisfy the resonance condition. In the
following, we estimate the overhead of obtaining the eigenvalue of an
eigenstate in the state evolution path through the QRT method. For
simplicity, we assume that the state evolution path is composed of a
sequence of ground states of the Hamiltonians \{$H_{k}$\}. For some
problems, the ground state eigenvalues of the Hamiltonians are already
known, then the parameters can be set directly in the QRT method. In
general, in order to obtain the ground state eigenvalue, we first guess a
range of the ground state eigenvalue, get a frequency range for the
resonance frequency, and discretize the frequency range into a number of
points, then apply the QRT method to obtain the ground state eigenvalue by
scanning these frequency points.

The QRT method performs quantum simulation of the dynamics, it follows the
same scaling laws as quantum simulation. There are two factors related to
the efficiency of this method in obtaining the ground state eigenvalue of a
Hamiltonian: ($1$) the number of experiments that need to be performed to
obtain the ground state eigenvalue in the scanned frequency range; ($2$) the
time needed to run the QRT method in an experiment. The total number of
experiments is equal to $ML$, where $M$ represents the number of frequency
points in the given frequency range, and $L$ represents the number of
experiments required to have a statistical estimation of the transition
probability $p_{k}\left( \omega \right) $ for a frequency $\omega $. To
determine the ground state eigenvalue within accuracy $\eta $, the number of
experiments $M$ required scales as $O\left( 1/\eta ^{2}\right) $. In
practice, one may be able to obtain better scaling than $O\left(
1/\eta^{2}\right) $, since one can zoom in on the resonances and keep a
lower density of frequency points elsewhere. The second factor, i.e. the
time needed to run the algorithm, is determined by the computational cost of
implementing the unitary operation $e^{-iH^{\left( k\right) }t_{k}}$, with
the evolution time scales as $t_{k}\sim \pi /(2cd_{k})$. The implementation
of $e^{-iH^{\left( k\right) }t_{k}}$ obeys the rules of implementing the
time evolution operator of the quantum system. By applying the quantum signal
processing algorithm, the query complexity scales as $O\left(
\sum_{i=1}^{ML}s\left\Vert H^{(k)}\right\Vert _{\max }t_{k}+\frac{\log
1/\epsilon }{\log \log 1/\epsilon }\right) $, where $\left\Vert
H^{(k)}\right\Vert $ is the norm of the algorithm Hamiltonian $H^{(k)}$, and
$s$ is the sparsity of $H^{(k)}$, and $\epsilon $ represents accuracy.

\section{Error analysis of implementing the RUS-POVM approach through the
QRT method}

In this section, we analyze errors in the implementation of
the RUS-POVM approach through the QRT method. We first show that the unitary
operation in Eq.~($3$) can be implemented through the QRT method within
certain accuracy under the secular approximation, then we estimate the
success probability of implementing the RUS-POVM approach through the QRT
method.

In the QRT method for implementing the $k$th step of the
RUS-POVM approach, the Hamiltonians $H_{k-1}$ and $H_{k}$ satisfy $%
H_{k-1}|\varphi _{j}^{\left( k-1\right) }\rangle =E_{j}^{\left( k-1\right)
}|\varphi _{j}^{\left( k-1\right) }\rangle $, and $H_{k}|\varphi
_{j}^{\left( k\right) }\rangle =E_{j}^{\left( k\right) }|\varphi
_{j}^{\left( k\right) }\rangle $, respectively, where $j=0,1,\cdots ,N-1$. %
For simplicity, let states $|\varphi ^{\left( k-1\right)
}\rangle =|\varphi _{0}^{\left( k-1\right)}\rangle $ and $|\varphi ^{\left(
k\right) }\rangle =|\varphi _{0}^{\left(k\right) }\rangle $, and $E^{\left(
k-1\right) }=E_{0}^{\left( k-1\right) }$ and $E^{\left( k\right)
}=E_{0}^{\left( k\right) }$. Let
\begin{equation}
H_{0}^{\left( k\right) }=-\frac{1}{2}\omega \sigma _{z}\otimes
I_{N}+H_{R}^{\left( k\right) },
\end{equation}%
then $H^{\left( k\right) }$ of Eq.~($4$) can be written as
\begin{equation}
H^{\left( k\right) }=H_{0}^{\left( k\right) }+cW^{(k)},
\end{equation}%
where $W^{(k)}=\sigma _{x}\otimes I_{N}$. The Hamiltonian $H_{0}^{\left(
k\right) }$ has eigenstates%
\begin{equation}
H_{0}^{\left( k\right) }|0\rangle |\varphi _{j}^{\left( k-1\right) }\rangle
=\left( \frac{-\omega }{2}\!+\!\alpha _{k}E_{j}^{(k-1)}\right) |0\rangle
|\varphi _{j}^{\left( k-1\right) }\rangle ,
\end{equation}%
and%
\begin{equation}
H_{0}^{\left( k\right) }|1\rangle |\varphi _{j}^{\left( k\right) }\rangle
=\left( \frac{\omega }{2}\!+\!E_{j}^{(k)}\right) |1\rangle |\varphi
_{j}^{\left( k\right) }\rangle .
\end{equation}%
In the $k$th step, the system is evolved from the state $|0\rangle |\varphi
_{0}^{\left( k-1\right) }\rangle $ to the state $|1\rangle |\varphi
_{0}^{\left( k\right) }\rangle $ under the resonance condition of $\alpha
_{k}E_{0}^{\left( k-1\right) }-E_{0}^{\left( k\right) }=\omega $. Errors in
the computation are introduced by excitations from the initial state to the
undesired eigenstates $|\varphi_{j}^{\left( k\right) }\rangle $~($j=1,\cdots ,N-1$) of $H_{k}$.

For an ancillary qubit coupled to a two-level system described by the
Hamiltonian of Eq.~($4$), according to the Rabi's formula~\cite{cohen}, the
transition probability from the excited state to the ground state of the
two-level system becomes higher as the transition frequency between the
two-level system gets closer to that of the ancillary qubit. In
the QRT method, a system with Hamiltonian $H_k$ that coupled to an ancillary
qubit contains the desired eigenstate $|\varphi_{0}^{\left( k\right) }\rangle $
and the undesired eigenstates $|\varphi_{j}^{\left( k\right) }\rangle $, ($j=1,\ldots ,N-1$), and they are orthogonal to each other. When the resonance condition is satisfied, the system is evolved from the initial state to the desired state through
quantum resonance transition, there is also leakage from the initial state
to the undesired states of the system. The desired state is decoupled from the undesired
states, therefore the undesired states can be grouped together and
represented effectively by a single state.

Based on the above analysis, the upper bound of the error of a step, that
is, the upper bound of the transition probability from the initial state to
the undesired states of the step, can be obtained by assuming
all the undesired eigenstates are degenerate at the eigenstate of the Hamiltonian $%
H_{k}$ that is closest to the desired eigenstate $|\varphi ^{\left( k\right)
}\rangle $. We denote the undesired state as $|\varphi _{\bot }^{\left(
k-1\right) }\rangle $ for the Hamiltonian $H_{k-1}$ and $|\varphi
_{\bot }^{\left( k\right) }\rangle $ for the Hamiltonian $H_{k}$. Then in
basis $\left\{ |0\rangle |\varphi ^{\left( k-1\right) }\rangle ,|0\rangle
|\varphi _{\bot }^{\left( k-1\right) }\rangle ,|1\rangle |\varphi ^{\left(
k\right) }\rangle ,|1\rangle |\varphi _{\bot }^{\left( k\right) }\rangle
\right\} $, the Hamiltonian ${H^{(k)}}$ can be written as:
\begin{equation}
{H^{(k)}=}\left(
\begin{array}{cccc}
E_{0} & 0 & c{d}_{k} & c\sqrt{1-{d}_{k}^{2}} \\
0 & E_{2} & c\sqrt{1-{d}_{k}^{2}} & c{d}_{k} \\
c{d}_{k} & c\sqrt{1-{d}_{k}^{2}} & E_{1} & 0 \\
c\sqrt{1-{d}_{k}^{2}} & c{d}_{k} & 0 & E_{3}%
\end{array}%
\right) ,
\end{equation}%
where $E_{0}=\frac{-\omega }{2}+\alpha _{k}E^{(k-1)}=\frac{1}{2}\omega
+E^{(k)}=E_{1}$ under the resonance condition $\alpha _{k}E^{\left(
k-1\right) }-E^{\left( k\right) }=\omega $, and $E_{2}=E_{0}+\Delta _{1}$, $%
E_{3}=E_{0}+\Delta _{2}$, $\Delta _{1}$ and $\Delta _{2}\gg c$ are
polynomial large. Then we have
\begin{equation}
H_{0}^{\left( k\right) }{=}\left(
\begin{array}{cccc}
E_{0} & 0 & 0 & 0 \\
0 & E_{2} & 0 & 0 \\
0 & 0 & E_{0} & 0 \\
0 & 0 & 0 & E_{3}%
\end{array}%
\right) ,
\end{equation}%
and
\begin{equation}
W{^{(k)}=}\left(
\begin{array}{cccc}
0 & 0 & {d}_{k} & \sqrt{1-{d}_{k}^{2}} \\
0 & 0 & \sqrt{1-{d}_{k}^{2}} & {d}_{k} \\
{d}_{k} & \sqrt{1-{d}_{k}^{2}} & 0 & {0} \\
\sqrt{1-{d}_{k}^{2}} & {d}_{k} & 0 & 0%
\end{array}%
\right) .
\end{equation}%
Let $|\psi \left( t\right) \rangle =c_{0}(t)e^{-iE_{0}t}|0\rangle |\varphi
^{\left( k-1\right) }\rangle +c_{1}(t)e^{-iE_{2}t}|0\rangle |\varphi _{\bot
}^{\left( k-1\right) }\rangle +c_{2}(t)e^{-iE_{0}t}|1\rangle |\varphi
^{\left( k\right) }\rangle +c_{3}(t)e^{-iE_{3}t}|1\rangle |\varphi _{\bot
}^{\left( k\right) }\rangle $, solving the Schr\"{o}dinger equation, we have
\begin{equation}
i\frac{d}{dt}c_{j}(t)=c\sum_{l}e^{i(E_{j}-E_{l})t}W_{jl}^{\left( k\right)
}c_{l}(t).
\end{equation}%
By setting the initial state as $|0\rangle |\varphi ^{\left( k-1\right)
}\rangle $, we have $c_{0}(0)=1$ and $c_{j}(0)=0$ for $j\neq 0$. The
coefficients of $c_{0}(t)$ and $c_{2}(t)$ are proportional to one, so they
oscillate slowly in time under the resonance condition, while the
coefficients $c_{1}(t)$ and $c_{3}(t)$ oscillate much more rapidly. By
applying the secular approximation~\cite{cohen1}, the rapid oscillating
terms are neglected since their contribution is negligible when integrated
over time. Then the above equations are simplified to%
\begin{eqnarray}
i\frac{d}{dt}c_{0}(t) &=&c{d}_{k}c_{2}(t)  \notag \\
i\frac{d}{dt}c_{2}(t) &=&c{d}_{k}c_{0}(t),
\end{eqnarray}%
and the Hamiltonian of the QRT method is spanned in basis $\left\{ |0\rangle
|\varphi ^{\left( k-1\right) }\rangle ,|1\rangle |\varphi ^{\left( k\right)
}\rangle \right\} $ in form of
\begin{equation}
{H^{(k)\prime }=}\left(
\begin{array}{cc}
\frac{1}{2}\omega +E^{(k)} & c{d_{k}} \\
c{d}_{k} & \frac{1}{2}\omega +E^{(k)}%
\end{array}%
\right) .
\end{equation}%
The corresponding time evolution operator is
\begin{equation}
U_{k}^{\prime }=\exp \left[ -iH^{\left( k\right) \prime }t\right] {=}\left(
\begin{array}{cc}
\cos \theta _{k} & -i\sin \theta _{k} \\
-i\sin \theta _{k} & \cos \theta _{k}%
\end{array}%
\right) ,
\end{equation}%
where $\theta _{k}=c{d}_{k}t$. The unitary operator $U_{k}^{\prime }$
transforms the initial state $|0\rangle |\varphi ^{\left( k-1\right)
}\rangle $ to an entangled state
\begin{equation}
U_{k}^{\prime }|0\rangle |\varphi ^{\left( k-1\right) }\rangle =\cos \theta
_{k}|0\rangle |\varphi ^{\left( k-1\right) }\rangle -i\sin \theta
_{k}|1\rangle |\varphi ^{\left( k\right) }\rangle .
\end{equation}%
Therefore under the secular approximation, the unitary operation in Eq.~($3$%
) of the RUS protocol can be implemented through the QRT method, and the
probability of the initial state $|0\rangle |\varphi ^{\left( k-1\right)
}\rangle $ being evolved to the state $|1\rangle |\varphi ^{\left( k\right)
}\rangle $ is $\sin ^{2}\left( cd_{k}t\right) $.

By setting $\Delta _{1}=\Delta _{2}=\Delta $ for simplicity, the Eq.~($13$)
can be solved analytically. In basis of $\left\{ |0\rangle |\varphi ^{\left(
k-1\right) }\rangle ,|0\rangle |\varphi _{\bot }^{\left( k-1\right) }\rangle
,|1\rangle |\varphi ^{\left( k\right) }\rangle ,|1\rangle |\varphi _{\bot
}^{\left( k\right) }\rangle \right\} $, let $|\psi _{i}\rangle =\left(
1,0,0,0\right) ^{\text{T}}$ be the initial state and $|\psi _{f}\rangle
=\left( 0,0,1,0\right) ^{\text{T}}$ be the final state, the transition
probability from the state $|\psi _{i}\rangle $ to the state $|\psi
_{f}\rangle $ is
\begin{eqnarray}
p_{if}^{(k)}\left( t\right) &=&\left\vert \langle \psi _{f}|U_{k}|\psi
_{i}\rangle \right\vert ^{2}=\left\vert \langle \psi _{f}|\exp \left[
-iH^{\left( k\right) }t\right] |\psi _{i}\rangle \right\vert ^{2}  \notag \\
&=&\sin ^{2}\theta _{k}\left[ \!\frac{\Delta ^{2}}{\Omega _{k}^{2}}\sin
^{2}\left( \frac{\Omega _{k}t}{2}\right) +\cos ^{2}\left( \frac{\Omega _{k}t%
}{2}\right) \!\right]
\end{eqnarray}%
where
\begin{equation}
\Omega _{k}=\sqrt{\Delta ^{2}+4c^{2}\left( 1-d_{k}^{2}\right) }.
\end{equation}%
Compared to the transition probability of $\sin ^{2}\theta _{k}$ from the
result of Eq.~($17$), the deviation is
\begin{eqnarray}
\sin ^{2}\theta _{k}-p_{if}^{(k)}\left( t\right) &=&\frac{4c^{2}\left( 1-{d}%
_{k}^{2}\right) }{4c^{2}\left( 1-{d}_{k}^{2}\right) +\Delta ^{2}}\sin
^{2}\theta _{k}\sin ^{2}\left( \frac{\Omega _{k}t}{2}\right)  \notag \\
&<&\frac{4c^{2}\left( 1-{d}_{k}^{2}\right) }{\Delta ^{2}}.
\end{eqnarray}%
This means that the unitary operation in Eq.~($3$) can be implemented as
Eq.~($17$) by using the QRT method within accuracy of $\frac{4c^{2}\left( 1-{%
d}_{k}^{2}\right) }{\Delta ^{2}}$.

In the following, by considering errors introduced in each
step of the computation, we estimate the success probability of implementing
the RUS-POVM approach through the QRT method. By ignoring the global phase,
the Hamiltonian of the $k$th step of the QRT method can be written as
\begin{equation}
{H^{(k)}=}\left(
\begin{array}{cccc}
0 & 0 & c{d}_{k} & c\sqrt{1-{d}_{k}^{2}} \\
0 & \Delta & c\sqrt{1-{d}_{k}^{2}} & c{d}_{k} \\
c{d}_{k} & c\sqrt{1-{d}_{k}^{2}} & 0 & 0 \\
c\sqrt{1-{d}_{k}^{2}} & c{d}_{k} & 0 & \Delta%
\end{array}%
\right) .
\end{equation}%
The measurement outcome on the ancillary qubit in the $k$th step of the
RUS-POVM approach is either in state $|0\rangle $ or $|1\rangle $. Then the
procedure of performing a unitary operation $U_{k}=\exp \left[ -iH^{\left(
k\right) }t_{k}\right] $ followed by a measurement on the ancillary qubit
can be described by operators $A_{0\mid k}=\langle 0|U_{k}|0\rangle _{a}$
and $A_{1\mid k}=\langle 1|U_{k}|0\rangle _{a}$ acting on the register $R$,
respectively, depending on the measurement outcome of $|0\rangle $ or $%
|1\rangle $. The operators $A_{0\mid k}$ and $A_{1\mid k}$ are effective
measurements associated with POVM elements $\left\{ A_{0\mid k}^{\dagger
}A_{0\mid k},A_{1\mid k}^{\dagger }A_{1\mid k}\right\} $.

By setting the initial state of the RUS-POVM computation as $|0\rangle
|\varphi ^{\left( 0\right) }\rangle $, in the basis of $A_{0\mid 1}$ of $%
\left\{ |\varphi ^{\left( 0\right) }\rangle ,|\varphi _{\bot }^{\left(
0\right) }\rangle \right\} $, the initial state is $\left( 1,0\right) _{0}^{%
\text{T}}$, and in the basis of $A_{1\mid m}$ of $\left\{ |\varphi ^{\left(
m\right) }\rangle ,|\varphi _{\bot }^{\left( m\right) }\rangle \right\} $,
the final state of the RUS-POVM process is $\left( 1,0\right) _{m}^{\text{T}%
} $ for a successful computation. Then the success probability of the
RUS-POVM process is:%
\begin{equation}
P_{\text{succ}}=\sum\limits_{k_{1},\ldots ,k_{m}}\left\vert \left(
\begin{array}{cc}
1 & 0%
\end{array}%
\right) _{m}A_{1\mid m}A_{0\mid m}^{k_{m}}\cdots A_{1\mid 1}A_{0\mid
1}^{k_{1}}\binom{1}{0}_{0}\right\vert ^{2}.
\end{equation}%
For simplicity we assume the overlaps ${d}_{k}=\left\vert \langle \varphi
^{(k-1)}|\varphi ^{(k)}\rangle \right\vert $ in each step are the same and
is denoted as $d$, and $t_{0}=\pi /(2c{d})$. It can be proved that the
success probability of the computation satisfies%
\begin{equation}
P_{\text{succ}}>\frac{1}{2}\left( 1+1/e\right) .
\end{equation}%
(See appendix A for detailed derivation).

To summarize, by implementing an $m$-step RUS-POVM approach through the QRT
method, the success probability of the method is lower-bounded by $\frac{1}{2%
}\left( 1+1/e\right) $, provided that the parameters $d$ and $\Delta $ are
polynomial large, and $c<\min \left\{ \frac{\Delta }{\sqrt{8m(1-{d}^{2})}}%
,\Delta \right\} $, the evolution time of each step is $t=\pi /(2c{d})$,
where $d$ is the overlap between the selected eigenstates of two adjacent
Hamiltonians ${H}_{k-1}$ and ${H}_{k}$, and $\Delta $ represents the energy
gap between the eigenstate $|\varphi^{(k)}\rangle $ and its nearest
neighboring eigenstates of ${H}_{k}$.

\section{Comparison of the RUS-POVM approach with the RUS procols and
measurement assisted methods}

The RUS-POVM approach is different from the existing RUS protocols and
measurement-assisted ancilla-driven quantum computing model. The major
difference is that in the RUS-POVM approach, the RUS protocol uses a unitary
operation that generates an entangled state between the ancillary qubit and
the working qubits, this entangled state bridges the two adjacent state in
the state evolution path. And because of this property, this approach can
perform multi-step quantum computation driven by POVM measurements on the
working qubits. In the following, we compare the RUS-POVM approach with the
previous RUS protocols and the ancilla-driven quantum computing model in
detail, respectively.

The RUS protocols in Refs.~\cite{wiebe,pae,boc,dong} and the RUS-POVM
approach share the core RUS principle, that is, repeat until a success is
obtained, but they are different in application and mechanism. The RUS
circuits in Refs.~\cite{wiebe,pae,boc} are mainly used in gate-level
compilation, they aim to approximate a target unitary~(typically a
single-qubit rotation) with minimal number of $T$ gates using a fixed
structure with one or more ancillas. They are fixed circuits with a single
success condition and incorporate Clifford corrections on failure. In Ref.~%
\cite{dong}, a success-or-draw framework was introduced to generalize the
RUS protocol to arbitrary probabilistic supermaps, it aims to realize an
operation of a given unitary operation, e.g., inverse, controlled
operations, etc. The success-or-draw method addresses the problem at the
level of quantum supermaps, i.e., transformations that map quantum
operations to quantum operations. It provides a universal construction that
takes any probabilistic supermap and converts it into a \textquotedblleft
success-or-draw\textquotedblright\ supermap, in which the identity operation
is explicitly constructed to act on the input state upon failure. This is
achieved by using a number of copies of the input unitary. If there exists a
probabilistic supermap transforming a $r$-dimensional unitary operation into
an arbitrary completely positive and trace-preserving map, then it is
possible to construct a success-or-draw supermap with $r$ copies of the
input unitary operation.

The RUS-POVM approach extends the idea of RUS to multi-step computation via
POVM measurements. It targets the construction of a quantum computation
process along a designed state evolution path. The computation consists of a
sequence of RUS steps, each step implements POVM measurements on the
register of the working qubits, and aims to obtain the desired state of the
step from an initial state. The POVM measurement is realized by applying a
unitary operation on the whole system followed by projective measurement on
the ancillary qubit, and the construction of the unitary operation depends
on the initial and final state of the step.

The mechanism of constructing the RUS protocol in the RUS-POVM approach is
different from the above methods. In this approach, an RUS protocol is
naturally constructed through a unitary operation that generates an
entangled state between the ancillary qubit and the working qubits, such
that upon measuring the ancilla, the working qubits either advance to the
next target state $|\varphi^{(k)}\rangle $~(success) or remain in its
current state $|\varphi^{(k-1)}\rangle $~(failure/draw). The number of times
the unitary operation in each step is queried depends only on the overlap
between two consecutive states of the state evolution path, not related to
the dimension of the unitary operation. This natural \textquotedblleft
draw\textquotedblright\ property is inherent to the RUS-POVM approach, not
an external addition. The physical realization of the RUS-POVM approach is
proposed via the QRT method. It demonstrates that the RUS protocol can be elevated from a gate-synthesis tool to a fundamental building block for quantum information processing.

Next, we compare the RUS-POVM approach with the ADQC model~\cite{bro},
which is a hybrid of the measurement-based quantum computing model and the circuit model. In ADQC, a fully controlled ancillary qubit is coupled to a computation register only
via a fixed unitary two-qubit interaction and then measured in suitable
bases, driving both single- and two-qubit operations on the register. There
is no direct manipulation on the register qubits.

The ADQC model and the RUS-POVM approach both harness a single ancillary
qubit to drive quantum computation through measurements, yet they differ in
their underlying principles and operational scopes. The original ADQC in
Ref.~\cite{bro} relies on a fixed maximally entangling interaction~(e.g., CZ
or CZ+SWAP) to implement deterministic, stepwise single- and two-qubit
quantum gates, with measurement outcomes compensated by Pauli operation
corrections. A subsequent extension of the ADQC model in Ref.~\cite{sha}
relaxes the requirement of maximal entangling power, allowing arbitrary
interaction strengths at the cost of probabilistic two-qubit gates;
universality is regained via RUS strategies and gate approximations, while
single-qubit operations remain deterministic. The RUS-POVM approach
generalizes the RUS concept to the entire computation path: each step
corresponds to a POVM measurement on the working register induced by a
unitary operation on the ancilla-register system and followed by a
projective measurement on the ancilla, and the process is repeated until a
successful outcome heralds the desired state of the step.

The method in the ADQC model that implements single- and two-qubit gates on
the quantum register can be used for realization of the unitary operation in
the RUS-POVM approach, one more ancillary qubit is needed to interact with
the ($n+1$) qubit system. As a result, the number of unitary operations will
also increase since one more controlled qubit is added, but compare with the
circuit in Fig.~$3$, it uses fewer quantum resources.

Compared to adiabatic quantum computing~\cite{AQC}, the RUS-POVM approach
has the advantage that only time-independent Hamiltonian evolution is
required in the computation. In comparison with the methods that employ
Zeno-like measurements~\cite{childs,boixo,poulin}, where frequent
measurements are performed to keep a quantum computer near the ground state
of a smoothly varying Hamiltonian, which require the overlap between
eigenstates of two adjacent Hamiltonians to be close to $1$ and performs
measurement on multiple qubits, our approach requires that the overlap
between two adjacent states of the state evolution path to be polynomial
large. It also avoids frequent measurements and measurement is performed
only on a single qubit. The technique of performing intermediate
measurements on a single qubit has been realized in ion-trap system~\cite%
{iontrap}.

\section{Conclusion}

In this work, we present a quantum computation approach that contains
intermediate measurements in the computation process. Simulating this
approach by deferring all the intermediate measurements to the end of the
circuit requires more resources. We demonstrate that by introducing
measurements in the intermediate stages of the circuit, both the number of
qubits and unitary operations can be reduced polynomially, and the
complexity of quantum circuits is simplified substantially, compared to the
case where all the measurements are deferred to the end of the circuit. A
method for implementing the approach is also provided based on quantum
resonant transitions. In the RUS-POVM approach, the system can jump from one
state to another following a given state evolution path. This provides
flexibility in constructing an efficient computation path, provided the
probability of success measurement in each step is polynomially large. How
to construct a physically implementable state evolution path for a specific
problem needs further study.

\begin{acknowledgements}
We thank Shuxian Chen for helpful discussions. This work was supported by the National Key Research and Development Program of China~(Grant No.~2023YFA1009103), Innovation Program for Quantum Science and Technology~(Grant No.~2021ZD0300804), the Guangdong Provincial Basic and Applied Basic Research Foundation~(Grant No.~2026B0303050003), and the Natural Science Basic Research Program of Shaanxi~(Program No.~2026JC-JCQN-001).
\end{acknowledgements}

\appendix

\section{Estimation of the success probability of implementing the RUS-POVM
approach via the QRT method}

We provide details for estimating the success probability of implementing
the RUS-POVM approach via the QRT method. By ignoring the global phase, the
Hamiltonian of the $k$th step of the approach is:
\begin{equation}
{H^{(k)}=}\left(
\begin{array}{cccc}
0 & 0 & c{d}_{k} & c\sqrt{1-{d}_{k}^{2}} \\
0 & \Delta & c\sqrt{1-{d}_{k}^{2}} & c{d}_{k} \\
c{d}_{k} & c\sqrt{1-{d}_{k}^{2}} & 0 & 0 \\
c\sqrt{1-{d}_{k}^{2}} & c{d}_{k} & 0 & \Delta%
\end{array}%
\right) .
\end{equation}%
Let $H_{d}$ be the Hadamard matrix,
\begin{equation}
{H_{d}=}\frac{1}{\sqrt{2}}\left(
\begin{array}{cc}
1 & 1 \\
1 & -1%
\end{array}%
\right) ,
\end{equation}%
and define $\widetilde{H}^{(k)}=\left( {H_{d}\otimes I_{2}}\right) {%
H^{(k)}\left( {H_{d}\otimes I_{2}}\right) }$, where $I_{2}$ is the
two-dimensional identity operator, then $\widetilde{H}^{(k)}$ is in form of
\begin{equation}
\widetilde{H}^{(k)}{=}\!\left( \!\!%
\begin{array}{cccc}
c{d}_{k} & c\sqrt{1\!-\!{d}_{k}^{2}} & 0 & 0 \\
c\sqrt{1\!-\!{d}_{k}^{2}} & c{d}_{k}\!+\!\Delta & 0 & 0 \\
0 & 0 & -c{d}_{k} & -c\sqrt{1\!-\!{d}_{k}^{2}} \\
0 & 0 & -c\sqrt{1\!-\!{d}_{k}^{2}} & -c{d}_{k}\!+\!\Delta%
\end{array}%
\!\!\right) .
\end{equation}%
We have
\begin{equation}
\exp \left[ -i\widetilde{H}^{(k)}t_{k}\right] =\left(
\begin{array}{cc}
U_{k}^{+} & 0 \\
0 & U_{k}^{-}%
\end{array}%
\right) =U_{k}^{+}\oplus U_{k}^{-}.
\end{equation}%
where
\begin{widetext}
\begin{eqnarray}
U_{k}^{+}{=}e^{-i(cd_{k}+\frac{\Delta }{2})t_{k}}\left(
\begin{array}{cc}
\cos \frac{\Omega _{k}t_{k}}{2}+i\frac{\Delta }{\Omega _{k}}\sin \frac{%
\Omega _{k}t_{k}}{2} & -i\frac{2c\sqrt{1-{d}_{k}^{2}}}{\Omega _{k}}\sin
\frac{\Omega _{k}t_{k}}{2} \\
-i\frac{2c\sqrt{1-{d}_{k}^{2}}}{\Omega _{k}}\sin \frac{\Omega _{k}t_{k}}{2}
& \cos \frac{\Omega _{k}t_{k}}{2}-i\frac{\Delta }{\Omega _{k}}\sin \frac{%
\Omega _{k}t_{k}}{2}%
\end{array}%
\right).
\end{eqnarray}
\begin{eqnarray}
U_{k}^{-}{=}e^{i(cd_{k}-\frac{\Delta }{2})t_{k}}\left(
\begin{array}{cc}
\cos \frac{\Omega _{k}t_{k}}{2}+i\frac{\Delta }{\Omega _{k}}\sin \frac{%
\Omega _{k}t_{k}}{2} & i\frac{2c\sqrt{1-{d}_{k}^{2}}}{\Omega _{k}}\sin \frac{%
\Omega _{k}t_{k}}{2} \\
i\frac{2c\sqrt{1-{d}_{k}^{2}}}{\Omega _{k}}\sin \frac{\Omega _{k}t_{k}}{2} &
\cos \frac{\Omega _{k}t_{k}}{2}-i\frac{\Delta }{\Omega _{k}}\sin \frac{%
\Omega _{k}t_{k}}{2}%
\end{array}%
\right).
\end{eqnarray}
\end{widetext}

In the $k$th step of the computation, the measurement outcome on the
ancillary qubit is either in state $|0\rangle $ or $|1\rangle $. Then the
procedure of performing a unitary operation $U_{k}=\exp \left[ -iH^{\left(
k\right) }t_{k}\right] $ followed by a measurement on the ancillary qubit
can be described by operators $A_{0\mid k}=\langle 0|U_{k}|0\rangle _{a}$
and $A_{1\mid k}=\langle 1|U_{k}|0\rangle _{a}$ applying on the register $R$%
, respectively, depending on the measurement outcome of state $|0\rangle $
or $|1\rangle $, the operators $A_{0\mid k}$ and $A_{1\mid k}$ are effective
measurements associated with POVM elements $\left\{ A_{0\mid k}^{\dagger
}A_{0\mid k},A_{1\mid k}^{\dagger }A_{1\mid k}\right\} $.

The unitary operation $U_{k}$ is
\begin{eqnarray}
U_{k} &=&\left( {H_{d}\otimes I_{2}}\right) \exp \left[ -i\widetilde{H}%
^{(k)}t_{k}\right] {\left( {H_{d}\otimes I_{2}}\right) }  \notag \\
&{=}&\frac{1}{2}\left(
\begin{array}{cc}
U_{k}^{+}+U_{k}^{-} & U_{k}^{+}-U_{k}^{-} \\
U_{k}^{+}-U_{k}^{-} & U_{k}^{+}+U_{k}^{-}%
\end{array}%
\right) .
\end{eqnarray}%
By denoting
\begin{equation}
c_{k}=\frac{2c\sqrt{1-d_{k}^{2}}}{\Omega _{k}},\quad s_{k}=\frac{\Delta }{%
\Omega _{k}},\quad T_{k}=cd_{k}t_{k},
\end{equation}%
the operators $A_{0\mid k}$ and $A_{1\mid k}$ are in the following form,
respectively,
\begin{widetext}
\begin{eqnarray}
A_{0\mid k} &=&\langle 0|U_{k}|0\rangle _{a}{=}\frac{1}{2}\left(
U_{k}^{+}+U_{k}^{-}\right)  \notag \\
&=&\left[
\begin{array}{cc}
\cos T_{k}\left( \cos \frac{\Omega _{k}t_{k}}{2}+is_{k}\sin \frac{\Omega
_{k}t_{k}}{2}\right) & -c_{k}\sin T_{k}\sin \frac{\Omega _{k}t_{k}}{2} \\
-c_{k}\sin T_{k}\sin \frac{\Omega _{k}t_{k}}{2} & \cos T_{k}\left( \cos
\frac{\Omega _{k}t_{k}}{2}-is_{k}\sin \frac{\Omega _{k}t_{k}}{2}\right)%
\end{array}%
\right]  \notag \\
&=&\cos T_{k}\cos \frac{\Omega _{k}t_{k}}{2}I_{2}-c_{k}\sin T_{k}\sin \frac{%
\Omega _{k}t_{k}}{2}\sigma _{x}+is_{k}\cos T_{k}\sin \frac{\Omega _{k}t_{k}}{%
2}\sigma _{z}.
\end{eqnarray}
and 
\begin{eqnarray}
A_{1\mid k} &=&\langle 1|U_{k}|0\rangle _{a}{=}\frac{1}{2}\left(
U_{k}^{+}-U_{k}^{-}\right)  \notag \\
&=&-\left[
\begin{array}{cc}
\sin T_{k}\left( \cos \frac{\Omega _{k}t_{k}}{2}+is_{k}\sin \frac{\Omega
_{k}t_{k}}{2}\right) & c_{k}\cos T_{k}\sin \frac{\Omega _{k}t_{k}}{2} \\
c_{k}\cos T_{k}\sin \frac{\Omega _{k}t_{k}}{2} & \sin T_{k}\left( \cos \frac{%
\Omega _{k}t_{k}}{2}-is_{k}\sin \frac{\Omega _{k}t_{k}}{2}\right)%
\end{array}%
\right]  \notag \\
&=&-\sin T_{k}\cos \frac{\Omega _{k}t_{k}}{2}I_{2}-c_{k}\cos T_{k}\sin \frac{%
\Omega _{k}t_{k}}{2}\sigma _{x}-is_{k}\sin T_{k}\sin \frac{\Omega _{k}t_{k}}{%
2}\sigma _{z}.
\end{eqnarray}
The POVM elements 
\begin{eqnarray}
A_{0\mid k}^{\dagger }A_{0\mid k}\!&=&\!\cos ^{2}T_{k}\!-\!c_{k}^{2}\cos 2T_{k}\sin
^{2}\frac{\Omega _{k}t_{k}}{2}\!-\!c_{k}\sin 2T_{k} \sin \frac{%
\Omega _{k}t_{k}}{2}\left(\!\!\cos \frac{\Omega _{k}t_{k}}{2}\sigma
_{x}\!+\!s_{k}\sin \frac{\Omega _{k}t_{k}}{2}\sigma _{y}\!\!\right)
\end{eqnarray}
\begin{eqnarray}
A_{1\mid k}^{\dagger }A_{1\mid k}\!&=&\!\sin ^{2}T_{k}+c_{k}^{2}\cos 2T_{k}\sin
^{2}\frac{\Omega _{k}t_{k}}{2}\!+\!c_{k}\sin 2T_{k}\sin \frac{\Omega _{k}t_{k}}{2%
}\left(\!\!\cos \frac{\Omega _{k}t_{k}}{2}\sigma _{x}\!+\!s_{k}\sin \frac{\Omega
_{k}t_{k}}{2}\sigma _{y}\!\!\right) .
\end{eqnarray}
\end{widetext} We also have
\begin{equation}
\lambda _{\min }(A_{0|k}^{\dagger }A_{0|k})=\cos ^{2}T_{k}-c_{k}^{2}\sin ^{2}%
\frac{\Omega _{k}t_{k}}{2}\cos 2T_{k}-f_{k}
\end{equation}%
\begin{equation}
\lambda _{\min }(A_{1|k}^{\dagger }A_{1|k})=\sin ^{2}T_{k}+c_{k}^{2}\sin ^{2}%
\frac{\Omega _{k}t_{k}}{2}\cos 2T_{k}-f_{k}
\end{equation}%
with
\begin{equation}
f_{k}=c_{k}\left\vert \sin 2T_{k}\sin \frac{\Omega _{k}t_{k}}{2}\right\vert
\sqrt{1-c_{k}^{2}\sin ^{2}\frac{\Omega _{k}t_{k}}{2}}
\end{equation}%
where $\lambda _{\min }(A_{0|k}^{\dagger }A_{0|k})$ and $\lambda _{\min
}(A_{1|k}^{\dagger }A_{1|k})$ are the minimum eigenvalue of the POVM
elements $A_{0|k}^{\dagger }A_{0|k}$ and $A_{1|k}^{\dagger }A_{1|k}$,
respectively. As a result
\begin{equation}
A_{a|k}^{\dagger n}A_{a|k}^{n}\geq \lambda _{\min }(A_{a|k}^{\dagger
}A_{a|k})A_{a|k}^{\dagger (n-1)}A_{a|k}^{n-1}\geq \lambda _{\min
}^{n}(A_{a|k}^{\dagger }A_{a|k}).
\end{equation}

By setting the initial state of the RUS-POVM process as $|0\rangle |\varphi
^{\left( 0\right) }\rangle $, in the basis of $A_{0\mid 1}$ of $\left\{
|\varphi ^{\left( 0\right) }\rangle ,|\varphi _{\bot }^{\left( 0\right)
}\rangle \right\} $, the initial state is $\left( 1,0\right) _{0}^{\text{T}}$%
, and in the basis of $A_{1\mid m}$ of $\left\{ |\varphi ^{\left( m\right)
}\rangle ,|\varphi _{\bot }^{\left( m\right) }\rangle \right\} $, the final
state of the RUS-POVM process is $\left( 1,0\right) _{m}^{\text{T}}$ for a
successful computation. Then the success probability of the RUS-POVM process
is:\begin{widetext}
\begin{eqnarray}
P_{\text{succ}} &=&\sum\limits_{k_{1},\ldots ,k_{m}}\left\vert \left(
\begin{array}{cc}
1 & 0%
\end{array}%
\right) _{m}A_{1\mid m}A_{0\mid m}^{k_{m}}\cdots A_{1\mid 1}A_{0\mid
1}^{k_{1}}\binom{1}{0}_{0}\right\vert ^{2}  \notag \\
&=&\sum\limits_{k_{1},\ldots ,k_{m}}\left[ \left(
\begin{array}{cc}
1 & 0%
\end{array}%
\right) _{m}A_{1\mid m}A_{0\mid m}^{k_{m}}\cdots A_{1\mid 1}A_{0\mid
1}^{k_{1}}\binom{1}{0}_{0}\right] ^{\dagger }\left(
\begin{array}{cc}
1 & 0%
\end{array}%
\right) _{m}A_{1\mid m}A_{0\mid m}^{k_{m}}\cdots A_{1\mid 1}A_{0\mid
1}^{k_{1}}\binom{1}{0}_{0}  \notag \\
&=&\sum\limits_{k_{1},\ldots ,k_{m}}\left( \!%
\begin{array}{cc}
1 & 0%
\end{array}%
\!\right) _{0}\left( A_{0\mid 1}^{\dagger }\right) \!\!^{k_{1}}\!A_{1\mid
1}^{\dagger }\cdots \left( A_{0\mid m}^{\dagger }\right) ^{k_{m}}\!A_{1\mid
m}^{\dagger }\binom{1}{0}_{m}\left( \!%
\begin{array}{cc}
1 & 0%
\end{array}%
\!\right) _{m}A_{1\mid m}A_{0\mid m}^{k_{m}}\cdots A_{1\mid 1}\!A_{0\mid
1}^{k_{1}}\binom{1}{0}_{0}  \notag \\
&=&\sum\limits_{k_{1},\ldots ,k_{m}}\left(
\begin{array}{cc}
1 & 0%
\end{array}%
\right) _{0}\left( A_{0\mid 1}^{\dagger }\right) \!\!^{k_{1}}A_{1\mid
1}^{\dagger }\cdots \left( A_{0\mid m}^{\dagger }\right) ^{k_{m}}A_{1\mid
m}^{\dagger }\frac{I_{2}+\sigma _{z}}{2}A_{1\mid m}A_{0\mid m}^{k_{m}}\cdots
A_{1\mid 1}A_{0\mid 1}^{k_{1}}\binom{1}{0}_{0}  \notag \\
&=&\frac{1}{2}\left( P_{0}+P_{1}\right) .
\end{eqnarray}
where 
\begin{eqnarray}
P_{0}=\sum\limits_{k_{1},\ldots ,k_{m}}\left(
\begin{array}{cc}
1 & 0%
\end{array}%
\right) _{0}\left( A_{0\mid 1}^{\dagger }\right) ^{k_{1}}A_{1\mid
1}^{\dagger }\cdots \left( A_{0\mid m}^{\dagger }\right) ^{k_{m}}A_{1\mid
m}^{\dagger }A_{1\mid m}A_{0\mid m}^{k_{m}}\cdots A_{1\mid 1}A_{0\mid
1}^{k_{1}}\binom{1}{0}_{0},
\end{eqnarray}
and 
\begin{eqnarray}
P_{1}=\sum\limits_{k_{1},\ldots ,k_{m}}\left(
\begin{array}{cc}
1 & 0%
\end{array}%
\right) _{0}\left( A_{0\mid 1}^{\dagger }\right) ^{k_{1}}A_{1\mid
1}^{\dagger }\cdots \left( A_{0\mid m}^{\dagger }\right) ^{k_{m}}A_{1\mid
m}^{\dagger }\sigma _{z}A_{1\mid m}A_{0\mid m}^{k_{m}}\cdots A_{1\mid
1}A_{0\mid 1}^{k_{1}}\binom{1}{0}_{0}.
\end{eqnarray}

By using the results of Eq.~(A$16$), we have 
\begin{eqnarray}
P_{0} &=&\sum\limits_{k_{1},\ldots ,k_{m}}\left(
\begin{array}{cc}
1 & 0%
\end{array}%
\right) _{0}\left( A_{0\mid 1}^{\dagger }\right) ^{k_{1}}A_{1\mid
1}^{\dagger }\cdots \left( A_{0\mid m}^{\dagger }\right) ^{k_{m}}A_{1\mid
m}^{\dagger }A_{1\mid m}A_{0\mid m}^{k_{m}}\cdots A_{1\mid 1}A_{0\mid
1}^{k_{1}}\binom{1}{0}_{0}  \notag \\
&\geq &\sum_{k_{1},\ldots ,k_{m}}\prod_{j=1}^{m}\lambda _{\min
}^{k_{j}}(A_{0|j}^{\dagger }A_{0|j})\lambda _{\min }(A_{1|j}^{\dagger
}A_{1|j})=\prod_{j=1}^{m}\frac{\lambda _{\min }(A_{1|j}^{\dagger }A_{1|j})}{%
1-\lambda _{\min }(A_{0|j}^{\dagger }A_{0|j})}.
\end{eqnarray}
Since 
\begin{eqnarray}
A_{0\mid k}^{\dagger }\sigma _{z}A_{0\mid k} &=&\left[ \cos T_{k}\cos \frac{%
\Omega _{k}t_{k}}{2}I_{2}-c_{k}\sin T_{k}\sin \frac{\Omega _{k}t_{k}}{2}%
\sigma _{x}-is_{k}\cos T_{k}\sin \frac{\Omega _{k}t_{k}}{2}\sigma _{z}\right]
\sigma _{z}  \notag \\
&&\left[ \cos T_{k}\cos \frac{\Omega _{k}t_{k}}{2}I_{2}-c_{k}\sin T_{k}\sin
\frac{\Omega _{k}t_{k}}{2}\sigma _{x}+is_{k}\cos T_{k}\sin \frac{\Omega
_{k}t_{k}}{2}\sigma _{z}\right]  \notag \\
&=&\left( \cos ^{2}T_{k}-c_{k}^{2}\sin ^{2}\frac{\Omega _{k}t_{k}}{2}\right)
\!\sigma _{z},
\end{eqnarray}
and 
\begin{eqnarray}
A_{1\mid k}^{\dagger }\sigma _{z}A_{1\mid k} &=&\left[ -\sin T_{k}\cos \frac{%
\Omega _{k}t_{k}}{2}I_{2}-c_{k}\cos T_{k}\sin \frac{\Omega _{k}t_{k}}{2}%
\sigma _{x}+is_{k}\sin T_{k}\sin \frac{\Omega _{k}t_{k}}{2}\sigma _{z}\right]
\sigma _{z}  \notag \\
&&\left[ -\sin T_{k}\cos \frac{\Omega _{k}t_{k}}{2}I_{2}-c_{k}\cos T_{k}\sin
\frac{\Omega _{k}t_{k}}{2}\sigma _{x}-is_{k}\sin T_{k}\sin \frac{\Omega
_{k}t_{k}}{2}\sigma _{z}\right]  \notag \\
&=&\left( \sin ^{2}T_{k}-c_{k}^{2}\sin ^{2}\frac{\Omega _{k}t_{k}}{2}\right)
\!\sigma _{z},
\end{eqnarray}
we have 
\begin{eqnarray}
P_{1} &=&\sum\limits_{k_{1},\ldots ,k_{m}}\left(
\begin{array}{cc}
1 & 0%
\end{array}%
\right) _{0}\left( A_{0\mid 1}^{\dagger }\right) ^{k_{1}}A_{1\mid
1}^{\dagger }\cdots \left( A_{0\mid m}^{\dagger }\right) ^{k_{m}}A_{1\mid
m}^{\dagger }\sigma _{z}A_{1\mid m}A_{0\mid m}^{k_{m}}\cdots A_{1\mid
1}A_{0\mid 1}^{k_{1}}\binom{1}{0}_{0}  \notag \\
&=&\prod_{k=1}^{m}\frac{\sin ^{2}T_{k}-c_{k}^{2}\sin ^{2}\frac{\Omega
_{k}t_{k}}{2}}{\sin ^{2}T_{k}+c_{k}^{2}\sin ^{2}\frac{\Omega _{k}t_{k}}{2}}
\geq \prod_{k=1}^{m}\frac{1-c_{k}^{2}-\cos ^{2}T_{k}}{1+c_{k}^{2}}.
\end{eqnarray}
\end{widetext}

We estimate the success probability of the computation by setting the
evolution time $t_{k}=t_{0|k}-\epsilon $ where $t_{0|k}=\pi /(2c{d}_{k})$
and $\epsilon >0$ is a small number, in this condition we have%
\begin{eqnarray}
P_{0} &\geq &\prod_{j=1}^{m}\frac{\lambda _{\min }(A_{1|j}^{\dagger }A_{1|j})%
}{1-\lambda _{\min }(A_{0|j}^{\dagger }A_{0|j})}  \notag \\
&=&\prod_{j=1}^{m}\frac{1-\tilde{f}_{j}}{1+\tilde{f}_{j}}%
=\prod_{j=1}^{m}(1-a_{j}\epsilon +o(\epsilon ^{2}))
\end{eqnarray}%
where we have denoted
\begin{eqnarray}
\tilde{f}_{k} &=&\frac{f_{k}}{\sin ^{2}T_{k}+c_{k}^{2}\sin ^{2}\frac{\Omega
_{k}t_{k}}{2}\cos 2T_{k}}  \notag \\
&=&\frac{2c_{k}\left\vert \sin T_{k}\sin \frac{\Omega _{k}t_{k}}{2}%
\right\vert \sqrt{1-c_{k}^{2}\sin ^{2}\frac{\Omega _{k}t_{k}}{2}}}{\sin
^{2}T_{k}+c_{k}^{2}\sin ^{2}\frac{\Omega _{k}t_{k}}{2}\cos 2T_{k}}|\cos
T_{k}|  \notag \\
&=&\frac{2c_{k}\left\vert \sin \frac{\Omega _{k}t_{0|k}}{2}\right\vert }{%
\sqrt{1\!-\!c_{k}^{2}\sin ^{2}\frac{\Omega _{k}t_{0|k}}{2}}}cd_{k}\epsilon
\!+\!o(\epsilon ^{2})  \notag \\
&:=&\frac{a_{k}}{2}\epsilon + o(\epsilon ^{2}).
\end{eqnarray}%
with
\begin{equation}
a_{k}=\frac{2cc_{k}d_{k}\left\vert \sin \frac{\Omega _{k}t_{0|k}}{2}%
\right\vert }{\sqrt{1-c_{k}^{2}\sin ^{2}\frac{\Omega _{k}t_{0|k}}{2}}}=\frac{%
8c^{2}d_{k}\sqrt{1-d_{k}^{2}}\left\vert \sin \frac{\Omega _{k}t_{0|k}}{2}%
\right\vert }{\Omega _{k}\sqrt{1-c_{k}^{2}\sin ^{2}\frac{\Omega _{k}t_{0|k}}{%
2}}}.
\end{equation}%
And
\begin{eqnarray}
P_{1} &\geq &\prod_{k=1}^{m}\frac{1-c_{k}^{2}-\cos ^{2}T_{k}}{1+c_{k}^{2}}
\notag \\
&\geq &\prod_{k=1}^{m}\left( 1-\frac{8c^{2}(1-d_{k}^{2})}{\Delta ^{2}}%
-\Omega _{k}^{2}c^{2}d_{k}^{2}\epsilon ^{2}\right) .
\end{eqnarray}

For simplicity we assume the overlaps ${d}_{k}$ in each step are the same
and is denoted as $d$, and $t_{0}=\pi /(2c{d})$, then we have
\begin{equation}
P_{0}\geq \left( 1-a\epsilon \right) ^{m}=1-ma\epsilon +O(\epsilon ^{2}),
\end{equation}%
where
\begin{equation}
a=\frac{8c^{2}{d}\sqrt{1-{d}^{2}}\left\vert \sin \frac{\Omega t_{0}}{2}%
\right\vert }{\Omega \sqrt{1-\frac{4c^{2}(1-{d}^{2})}{\Omega ^{2}}\sin ^{2}%
\frac{\Omega t_{0}}{2}}}.
\end{equation}%
And
\begin{eqnarray}
P_{1} &\geq &\left[ 1-\frac{8c^{2}(1-d^{2})}{\Delta ^{2}}-\Omega
^{2}c^{2}d^{2}\epsilon ^{2}\right] ^{m}  \notag \\
&=&\left[ 1-\frac{8c^{2}(1-d^{2})}{\Delta ^{2}}\right] ^{m}+O(\epsilon ^{2}).
\end{eqnarray}%
By setting the parameter $c$ such that $\frac{8(1-{d}^{2})}{\Delta ^{2}}%
c^{2}<\frac{1}{m}$, we have $P_{1}>1/e$. Then the success probability of the
computation satisfies%
\begin{equation}
P_{\text{succ}}=\frac{1}{2}\left( P_{0}+P_{1}\right) >\frac{1}{2}\left(
1+1/e\right) .
\end{equation}

\section{Example for the RUS-POVM approach}

We present an example for the RUS-POVM approach, focusing on quantum resource saving compared to the case where all the intermediate measurements are deferred to the end of the computation. The Hamiltonian evolution path of the example should satisfy the two conditions of the QRT method: $i$) the energy gap between the ground and the first excited states of each Hamiltonian, and $ii$) the overlap between the selected eigenstates of any two adjacent Hamiltonians are not exponentially small.

The Hamiltonian evolution path of the example is $H_{0}\rightarrow
H_{1}\rightarrow \dots \rightarrow H_{m}$, and $|V_{k}\rangle =\frac{1}{\sqrt{2^{k}}}\sum_{i=0}^{2^{k}-1}|\psi _{i}\rangle $ is an eigenstate of the Hamiltonian $H_{k}$ ($k=0,1,\cdots ,m$) with eigenvalue of one, and $\langle
\psi _{i}|\psi _{j}\rangle =\delta _{ij}$. Suppose the energy gap between
the state $|V_{k}\rangle $ and the nearest eigenstate of the Hamiltonian $H_{k}$ is polynomial large. The overlap between eigenstates $|V_{k-1}\rangle
$ and $|V_{k}\rangle $ of two adjacent Hamiltonians is $d=1/\sqrt{2}$, which
is also polynomial large. Therefore both conditions of the QRT method are
satisfied, and the computation can be run efficiently. By setting the
parameters $\omega =1$ and $\alpha _{k}=2$ in each step of QRT method such
that the resonance transition condition is satisfied, the evolution from the
initial state $|V_{0}\rangle $ to the final state $|V_{m}\rangle $ can be
run efficiently in $m$ steps. If one starts from the initial state $|V_{0}\rangle $ and evolves it directly to the state $|V_{m}\rangle $ in one step, the evolution time scales as $O\left( \sqrt{2^{m}}\right) $, since the overlap between the states $|V_{0}\rangle $ and $|V_{m}\rangle $ is $1/\sqrt{2^{m}}$. There are examples in which we can construct Hamiltonian evolution paths that have the properties as the above example, e.g., the Google matrix~\cite{pagerank}.

We consider a $10$-qubit system that requires $m=10$ steps in the RUS-POVM approach. The number of the unitary operations $U_{k}$ queried in the
approach is proportional to $\sum_{k=1}^{10}n_{k}$. To implement the
RUS-POVM approach for solving the problem using the QRT method, we set $c=0.02$, and the evolution time $t=50$, then the success probability of each
step is $p\approx \sin ^{2}(cdt)=0.42$. Therefore we can set $n_{k}=3$ in
each step of the RUS-POVM approach for solving this problem, and the unitary
operation $U_{k}=e^{-iH^{(k)}t}$ is queried $3$ times on average in each
step, the total number of times for querying the unitary operations is $30$.

In solving this problem by applying the circuit in Fig.~$3$ where all the intermediate measurements are deferred to the end of the circuit, the total
success probability of the computation is $\prod\limits_{k=1}^{m}\left[
1-(1-p)^{n_{k}}\right] =\left[ 1-(1-p)^{n_{k}}\right] ^{10}$. By setting the
number of ancillary qubits in each step as $n_{k}=3$, the success
probability is $0.11$. In simulating the $k$th step of the RUS-POVM approach
with $n_{k}=3$, the circuit contains one unitary operation $U_{k}$, a
single-qubit-controlled $U_{k}$ operation $C$-$U_{k}$ and a
two-qubit-controlled $U_{k}$ operation $C^{2}$-$U_{k}$. By adding one more
ancilla qubit, the $C^{2}$-$U_{k}$ operation can be implemented by two
Toffoli gates and a $C$-$U_{k}$ operation. Note this additional ancillary
qubit can be used in the following steps of the circuit. Then in the case of
$n_{k}=3$, by deferring all the intermediate measurements to the end of the
circuit, we need $4$ ancillary qubits, and one unitary operation $U_{k}$, two $C$-$U_{k}$ operations, and two Toffoli gates for simulating the $k$-th step of
the RUS-POVM approach. Therefore to simulate the RUS-POVM approach for
solving this problem, it requires a total number of $31$ ancillary qubits,
and $10$ $U_{k}$ operations, $20$ $C$-$U_{k}$ operations, and $20$ Toffoli
gates.

By setting $n_{k}=8$, we need to implement $80$ unitary operations $U_{k}$ in the RUS-POVM approach, while the success probability of the computation
in the circuit of Fig.~$3$ becomes $0.88$. We need $8$ ancillary qubits for
simulating the $k$th step of the RUS-POVM approach by using the circuit of
Fig.~$3$, the unitary operations are \{$U_{k}$, $C$-$U_{k}$, $C^{2}$-$U_{k}$,\ldots , $C^{7}$-$U_{k}$\} in the circuit can be implemented through one
unitary operation $U_{k}$, $7$ $C$-$U_{k}$ operations and $42$ Toffoli gates
by adding $7$ more ancillary qubits. Therefore it requires a total number of
$87$ ancillary qubits, and $10$ $U_{k}$ operations, $70$ $C$-$U_{k}$
operations, and $420$ Toffoli gates for simulating the RUS-POVM approach for
solving this problem.


\begin{thebibliography}{99}
\bibitem{QSP} G.~H. Low and I.~L. Chuang, Optimal Hamiltonian Simulation by
Quantum Signal Processing, Phys. Rev. Lett., \textbf{118}, 010501~(2017).

\bibitem{mea} R. Raussendorf and H.~J. Briegel, A One-Way Quantum Computer,
Phys. Rev. Lett. \textbf{86}, 5188~(2001).

\bibitem{bro} J. Anders, D.~K.~L. Oi, E. Kashefi, D.~E. Browne, and E. Andersson,
Ancilla-driven universal quantum computation, Phys. Rev. A \textbf{82},
020301~(2010).

\bibitem{sha} K.~H. Shah and D.~K.~L. Oi, Ancilla driven quantum computation
with arbitrary entangling strength, \textit{eprint}: arXiv:1303.2066.

\bibitem{poulin} D. Poulin, A.~Y. Kitaev, D.~S. Steiger, M.~B. Hastings and
M. Troyer, Quantum Algorithm for Spectral Measurement with a Lower Gate
Count, Phys. Rev. Lett. 121, 010501~(2018).

\bibitem{parker} S. Parker and M.~B. Plenio, Efficient Factorization with a
Single Pure Qubit and logN Mixed Qubits, Phys. Rev. Lett. \textbf{85},
3049~(2000).

\bibitem{iontrap} T. Monz, D. Nigg, E. A. Martinez, M. F. Brandl, P.
Schindler, R. Rines, S. X. Wang, I. L. Chuang, and R. Blatt, Realization of
a scalable Shor algorithm, Science 351, 1068 (2016).

\bibitem{wiebe} N. Wiebe and V. Kliuchnikov, Floating point representations
in quantum circuit synthesis, New J. Phys. \textbf{15}, 093041~(2013).

\bibitem{pae} A. Paetznick and K.~M. Svore, Repeat-Until-Success:
Non-deterministic decomposition of single-qubit unitaries, Quantum Inf.
Comput. \textbf{14}, 1277~(2014).

\bibitem{boc} A. Bocharov, M. Roetteler and K.~M. Svore, Efficient Synthesis
of Universal Repeat-Until-Success Quantum Circuits, Phys. Rev. Lett. \textbf{%
114}, 080502~(2015).

\bibitem{dong} Q. Dong, M. Quintino, A. Soeda and M. Murao, Success-or-Draw:
A Strategy Allowing Repeat-Until-Success in Quantum Computation, Phys. Rev.
Lett. \textbf{126}, 150504~(2021).

\bibitem{AQC} E. Farhi, J. Goldstone, S. Gutmann, and M. Sipser, Quantum
computation by adiabatic evolution, arXiv:quant-ph/0001106.

\bibitem{childs} A.~M. Childs, E. Deotto, E. Farhi, J. Goldstone, S. Gutmann
and A.~J. Landahl, Quantum search by measurement, Phys. Rev. A \textbf{66},
032314~(2002).

\bibitem{boixo} S. Boixo, E. Knill and R. Somma, Eigenpath traversal by
phase randomization, Quant. inf. comp. \textbf{9}, 0833~(2009).

\bibitem{aharanov} D. Aharanov and A. Ta-Shma, Adiabatic quantum state generation, SIAM J. Comput. \textbf{37}, 47~(2007).

\bibitem{sch} G. Schaller, Adiabatic preparation without quantum phase
transitions, Phys. Rev. A \textbf{78}, 032328~(2008).

\bibitem{nc} M.~A. Nielsen and I.~L. Chuang, Quantum computation and quantum
information~(Cambridge Univ. Press, Cambridge, England, 2000).

\bibitem{barenco} A. Barenco, C.~H. Bennett, R. Cleve, D.~P. Di Vincenzo, N.
Margolus, P. Shor, T. Sleator, J.~A. Smolin and H. Weinfurter, Elementary
gates for quantum computation, Phys. Rev. A \textbf{52}, 3457~(1995).

\bibitem{whf1} H. Wang, S. Ashhab and Franco Nori, Quantum algorithm for
obtaining the energy spectrum of a physical system, Phys. Rev. A \textbf{85}%
, 062304~(2012).

\bibitem{whf2} Z. Li, X. Liu, H. Wang, S. Ashhab, J. Cui, H. Chen, X. Peng,
and J. Du, Quantum Simulation of Resonant Transitions for Solving the
Eigenproblem of an Effective Water Hamiltonian, Phys. Rev. Lett. \textbf{122}%
, 090504~(2019).

\bibitem{wy} H. Wang and S. Yu, Quantum algorithm for preparing the ground
state of a physical system through multi-step quantum resonant transitions,
Quant. Inf. Proc., \textbf{20}, 40~(2021).

\bibitem{cohen} C. Cohen-Tannoudji, B. Diu, and F. Lalo\"{e}, \emph{Quantum
Mechanics} Vol.~1, p414, (Wiley-Interscience Publication 1977).

\bibitem{cohen1} C. Cohen-Tannoudji, B. Diu, and F. Lalo\"{e}, \emph{Quantum
Mechanics} Vol.~2, p1341, (Wiley-Interscience Publication 1977).

\bibitem{pagerank} C. Wang, H. Wang and H. Xiang, Quantum algorithm for PageRank computation through multistep quantum resonant transitions, J. Comput. Appl. Math. \textbf{491}, 118059~(2027).
\end{thebibliography}
\end{document}